# The EUCLID NISP grisms flight models performance

Anne Costille[*,a], A. Caillat[a], C. Rossin[b], S. Pascal[a], P. Sanchez[a], R. Barette[a], P. Laurent[a], B. Foulon[a], C. Pariès[a]

[a]Aix Marseille Univ, CNRS, CNES, LAM UMR 7326, 13388, Marseille, France, [b]Observatoire de Physique du Globe de Clermont-Ferrand, Campus Universitaire des Cézeaux, 4 Avenue Blaise Pascal, TSA 60026 - CS 60026, 63178 Aubière Cedex

**ABSTRACT**

ESA EUCLID mission will be launched in 2020 to understand the nature of the dark energy responsible of the accelerated expansion of the Universe and to map the geometry of the dark matter. The map will investigate the distance-redshift relationship and the evolution of cosmic structures thanks to two instruments: the NISP and the VIS. The NISP (Near Infrared Spectro-Photometer) is operating in the near-IR spectral range (0.9-2μm) with two observing modes: the photometric mode for the acquisition of images with broad band filters, and the spectroscopic mode for the acquisition of slitless dispersed images on the detectors. The spectroscopic mode uses four low resolution grisms to cover two spectral ranges: three "red" grisms for 1250-1850nm range, with three different orientations, and one "blue" grism for 920-1300nm range. The NISP grisms are complex optical components combining four main optical functions: a grism function (dispersion without beam deviation of the first diffracted order) done by the grating on the prism hypotenuse, a spectral filter done by a multilayer filter deposited on the first face of the prism to select the spectral bandpass, a focus function done by the curved filter face of the prism (curvature radius of 10m) and a spectral wavefront correction done by the grating which grooves paths are nor parallel, neither straight. The development of these components have been started since 10 years at the Laboratoire d'Astrophysique de Marseille (LAM) and was linked to the project phases: prototypes have been developed to demonstrate the feasibility, then engineering and qualification models to validate the optical and mechanical performance of the component, finally the flight models have been manufactured and tested and will be installed on NISP instrument. In this paper, we present the optical performance of the four EUCLID NISP grisms flight models characterized at LAM: wavefront error, spectral transmission and grating groove profiles. The test devices and the methods developed for the characterization of these specific optical components are described. The analysis of the test results have shown that the grisms flight models for NISP are within specifications with an efficiency better than 70% on the spectral bandpass and a wavefront error on surfaces better than 30nm RMS. The components have withstood vibration qualification level up to 11.6g RMS in random test and vacuum cryogenics test down to 130K with measurement of optical quality in transmission. The EUCLID grisms flight models have been delivered to NISP project in November 2017 after the test campaign done at LAM that has demonstrated the compliance to the specifications.

## 1. INTRODUCTION

EUCLID mission[1] has been selected by ESA in 2012 in the context of the Cosmic Vision program to study the nature of dark energy and dark matter. The mission is designed to map the geometry of the dark Universe by investigating the distance-redshift relationship and the evolution of cosmic structures thanks to two scientific instruments: the Near Infrared Spectroscopic Photometer (NISP)[2] and the Visible instrument (VIS)[3]. The NISP channel of Euclid is dedicated to measure the redshift of millions of galaxies and to analyze their spatial distribution in the Universe. NISP works with both photometric and spectroscopic modes by switching between broadband filters and grisms, mounted on two rotating wheels, to acquire data of the same field. The spectroscopic mode acquires dispersed images on the detector without a slit by using four different grisms mounted on a wheel.

*anne.costille@lam.fr; phone +33491055978; fax; +33491621190, www.lam.fr

The grisms designed for NISP are complex optical and mechanical components that have been studied deeply during phase A and B of NISP project[4] through the development of several prototypes. The project is now entering its final phase as the integration of the complete NISP instrument is started at Laboratoire d'Astrophysique de Marseille (LAM). Concerning the grisms, the Engineering and Qualification Model (EQM) and the four Flight Models (FM) have been delivered to the NISP grisms Wheel respectively at end of 2016 and end of 2017 and have been integrated and tested on the wheel in Spring 2018. We present in this paper the results of the performance analysis of the grisms FM, which will be used in NISP instrument. After a small recall of the design of the components, we present the optical performance of the four EUCLID NISP grisms flight models characterized at LAM: wavefront error, spectral transmission and grating groove profiles. We compare the results obtained on the four FM. We show also the behavior of the components during vibration tests and the results from the vacuum cryogenics test down to 130K with measurement of optical quality in transmission.

## 2. EUCLID GRISM DESCRIPTION

### 2.1 Euclid NISP grisms overview

The Euclid NISP grisms are critical parts of the NISP instrument as they are complex optical and mechanical components. In NISP, there are four grisms mounted on a wheel:

- The grism NI-GSU-FM-RGS000: a "red grism" with a 2.145° prism angle transmitting in spectral band [1.25-1.85µm] and oriented at 0° to obtain a spectrum vertical on NISP detector,
- The grism NI-GSU-FM-RGS180: a "red grism" with a 2.145° prism angle transmitting in spectral band [1.25-1.85µm] and oriented at 180° to obtain a spectrum vertical on NISP detector but with the opposite direction compared to the 0° orientation,
- The grism NI-GSU-FM-RGS270: a "red grism" with a 2.145° prism angle transmitting in spectral band [1.25-1.85µm] and oriented at 270° to obtain a spectrum horizontal on NISP detector,
- The grism NI-GSU-FM-BGS000: a "blue grism" with a 1.77° prism angle transmitting in spectral band [0.92-1.3 µm] and oriented at 0° to obtain a spectrum vertical on NISP detector.

Each grism is made of two parts, as shown in Figure 1:
- The optical element i.e. the grism itself,
- The mechanical mount, which maintains the grism integrity in cryogenic environment on the wheel (130K) and during the spacecraft launch. In addition, a baffle is fixed on the mount to limit scattered light in NISP instrument due to the component or any stray light coming from the telescope.

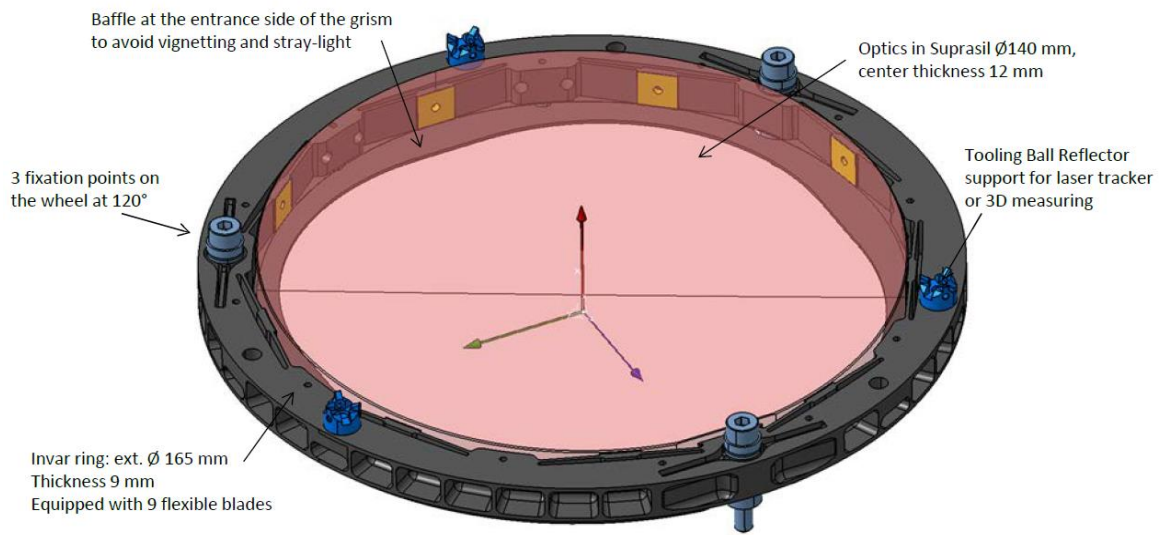

*Figure 1. CAD representation of EUCLID NISP RGS000. The yellow squares indicate the gluing areas.*

### 2.2 Euclid NISP grism optical part description

The optical part of NISP grism i.e. the grism itself, combines four optical functions in one component, which are represented in Figure 2:
- A grism in Suprasil 3001 made of a grating engraved on the prism hypotenuse to make the light undeviated at a chosen wavelength. In addition, a spectral wavefront correction is done by the curvature of the grating grooves,
- A spectral filter done by a multilayer filter deposited on the first surface of the prism,
- A focus function done by the curvature of the first surface of the prism.

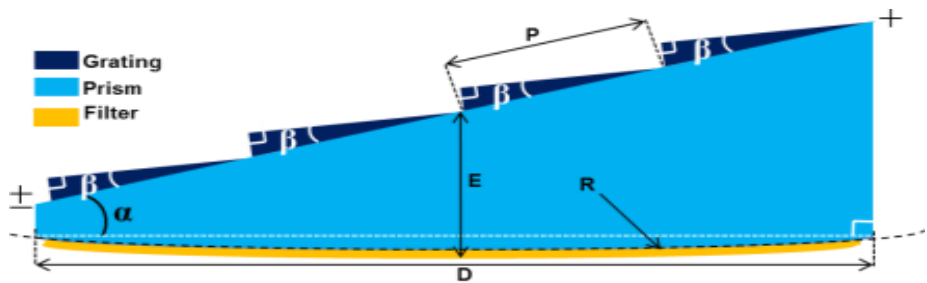

*Figure 2: Scheme of the optical part.*

The complete description and specifications of the FM grisms can be found in [4]. The main optical specifications of the grisms for NISP are:
- A spectral band transmission in order 1 better than 65% on the defined band passes,
- A spectral band transmission in order 0 better in total of 1% for calibration purpose;
- A large blocking band with a transmission lower than 5.e-4 on [400-550]nm and [920-2500]nm and lower than 2e-2 on [550-920]nm,

- A high quality of the optical surface with a RMS value of the grating surface lower than 15nm RMS and a filter surface lower than 10nm RMS before manufacturing of the filter and gluing into the mechanical mount,
- The manufacturing of the curved filter face with a curvature at +/-0.5fr from the nominal values. RGS000 and RGS180 have a designed curvature of -9631.06m, RGS270 of -9272.898m and BGS000 of -9906.874m,
- The transmitted WaveFront Error (WFE) of the component after gluing and filter manufacturing better than 15nm RMS. In addition, the bending due to the filter deposition should be lower than 5fr on each optical surface (filter and grating surfaces),
- The mean groove frequency equals to 13.75 grooves/mm for the "red" gratings and 15.1 grooves/mm for the blue grating.

The specificities of EUCLID NISP grisms are a grating with a low groove density and a small groove angle which is not usual for classical gratings manufacturing processes. In addition the grooves are not straight but curved to compensate for the focus chromatic aberrations of NISP instrument. The NISP detector is tilted with respect to the optical axis and the groove shape allows the focalization of the spectra on NISP detectors according to this tilt. The first surface of the grating is a convex surface with about -10m focal length, which is also quite difficult to achieve at the accuracy required by the tolerance analysis of NISP optical design. During phase B of NISP project, several R&D programs funded by the Centre National d'Etudes Spatiales (CNES) and the LAM have been done to develop the NISP gratings and results are presented in [4]. This R&D program has allowed to identify SILIOS Technologies Company as the manufacturer of the grating and to demonstrate the feasibility on prototypes. SILIOS has then been selected to manufacture the flight models for NISP instrument.

### 2.3 Euclid NISP grism mechanical part description

The optical part of the grisms is glued in a mechanical Invar M93 ring through 9 flexible blades that compensate the small CTE difference between Suprasil 3001 and Invar. The complete design of the mechanical mount and the analysis of its thermal and mechanical behavior have been done by LAM. Details of these analyses are presented in [5]. The gluing of the component into the mechanical mount is done by Winlight System Company and has been qualified during phase B of the project. In order to ensure a glue pads thickness of 100±10μm, the internal diameter of the mount is machined after measurement of the manufactured prism diameter. After manufacturing and tests at LAM, the mount is fixed on the Invar grism wheel thanks to three M4 screws. Three blades enable to minimize in the optical element stresses due to thermal differences (from 300K to 100K) and to limit interface defaults that deform the component at the wheel interface. The length and the thickness of the two stages of flexures are optimized to ensure the integrity of the assembly during thermal cycling and vibrations. The mount is designed to withstand design limit load of 60g with a first global resonance higher than 400Hz on a rigid structure.

### 2.4 EUCLID NISP grism manufacturing and test strategy

The complete manufacturing process and test sequence done on the NISP FM grisms is fully described in [6]. We recall in this section the main important points concerning the manufacturing and test strategy for the grisms that were established thanks to the development of prototypes and the Engineering and Qualification Model (EQM) developed at the beginning of phase D of the project. The manufacturing process of the optical part is quite complex as several manufacturers operate on the optical component one after the other. The complete manufacturing process follows the sequence below:
1. Manufacturing of a parallel plate of 150mm delivered by TRIOPTICS [7];
2. Manufacturing of the grating on one side of the parallel plate by SILIOS Technologies [8];
3. Manufacturing of the prism with the convex filter face opposite to the grating by WINLIGHT Optics [9];
4. Manufacturing of the filter done by Optics BALZERS Jena [10];
5. In parallel, manufacturing of the mechanical mount done by Alsyom;
6. Alignment and gluing of the optical part in the mechanical mount done by WINLIGHT System.

The test campaign of the grisms allows verifying the performance and the specifications of the components before the delivery to the project. All performance tests are under LAM responsibility. A complete test strategy has been elaborated to validate the performance of the components and is described in details in [6] with the test tools developed specifically for the characterization of the grisms. We recall here the main steps of the test campaign:

- Optical performance characterization before the gluing of the component into the mechanical mount:
    - We validate the transmission of the component into the bandpass with a spectrophotometer specially adapted to measure the grism on 90mm aperture[6]. The measurement is done at the end of the manufacturing process after the grating and the filter manufacturing;
    - We measure the groove profile of the grating with an interferential microscope Wyko NT9100 to verify the grating shape after manufacturing;
    - We measure the Surface Form Error (SFE) of the filter and grating surfaces thanks to a phase-shifting Fizeau interferometer working at 633nm. A special configuration has been developed to be able to measure the curved surface and to compensate for the large radius of curvature. Variation of the curved surface after the filter deposition is followed. The grating WFE function is directly measured in reflection and followed along the whole manufacturing process. The complete strategy for the SFE measurement of the grisms is fully described in [6];
- Focus measurement at 130K on the final component: the goal of this test is to verify that the global focus of the component does not vary between room temperature (300K) and operational temperature (130K). The measurement is done with a through focus method during a thermal vacuum cycling of the component. The best focus position at 130K is compared with the one measured at room temperature, the difference between the two positions should be lower than 100µm;
- Acceptance level vibrations of the components with sine acceptance level (X axis 30.5g, Y and Z axes 22.3g) and random test (11.6g RMS in X, 10.6g RMS in Y and 9.9g RMS in Z) with notches at the resonant frequencies of the grism;
- Metrology measurement of the assembled component on a Coordinates Measurement Machine (CMM) to measure the position of the optical component with respect to the mechanical mount. This measurement is done to measure with accuracy the localization of the center of the grating to align precisely the component onto the Grism Wheel Assembly.

The manufacturing of the four grisms FM has been started in early 2016 as the manufacturing time for one component is quite long. The manufacturing of a full component (optical part, mechanical part, gluing and test) lasts 9 months per component. The four FM for NISP have been delivered to the project at the end of 2017. These components are now fully assembled and aligned onto the grism wheel. Figure 3 presents a picture of each FM delivered to NISP project after the final inspection of the component. We can see on these pictures that the "red" grisms are very similar. The name identification of each component is engraved onto the mechanical mount to distinguish the components.

**NI-GSU-FM-RGS000**   **NI-GSU-FM-RGS180**   **NI-GSU-FM-RGS270**   **NI-GSU-FM-BGS000**

**NISP grism FM filter face**

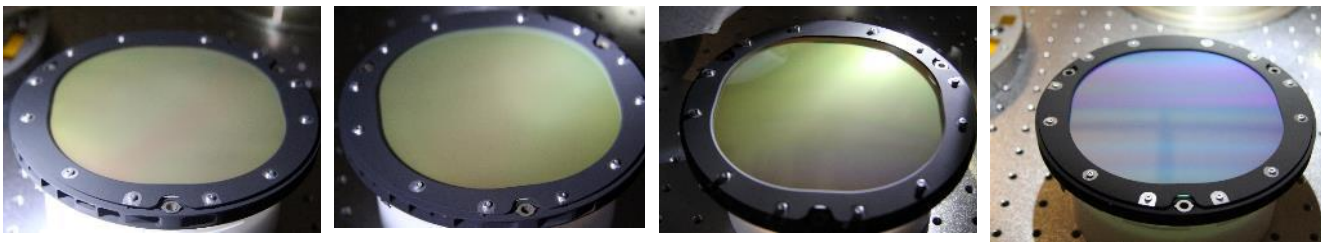

**NISP grism FM grating face**

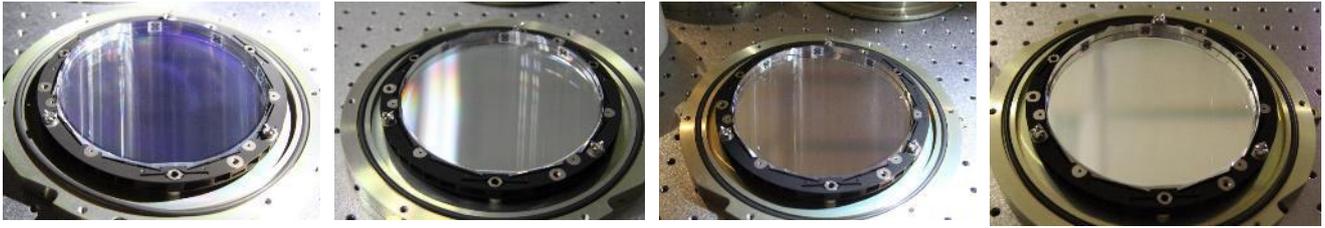

*Figure 3. The NISP grism flight models before delivery to NISP project. For each component, the filter and the grating surfaces are shown.*

## 3. OPTICAL PERFORMANCE OF NISP GRISMS FLIGHT MODELS

We present in this section the optical performance measured on each FM for NISP. The measurement of the optical performance is done with the test setups described in [6] according to the test plan of the grism components elaborated at the beginning of the project.

### 3.1 Groove profile measurement

The validation of the performance starts with the verification of the groove profile with the specified design. In particular, we check the groove height of each grating and the groove width. Figure 4 presents the grating groove profile measured on the four grisms FM. The measurements done on the three red gratings demonstrate the very good repeatability of the manufacturing process concerning the groove dimensions: period and height. The profile of the blue grating is provided for information but its period and height is different from the red gratings. We can see also the discretization of the groove slopes in 16 small steps due to the manufacturing process proposed by SILIOS Technologies[11], which uses several photolithographic masks and etching phases. The little peaks observed at certain steps are due to mis-alignment errors between the masks but have a very few impact on the overall performance of the grating.

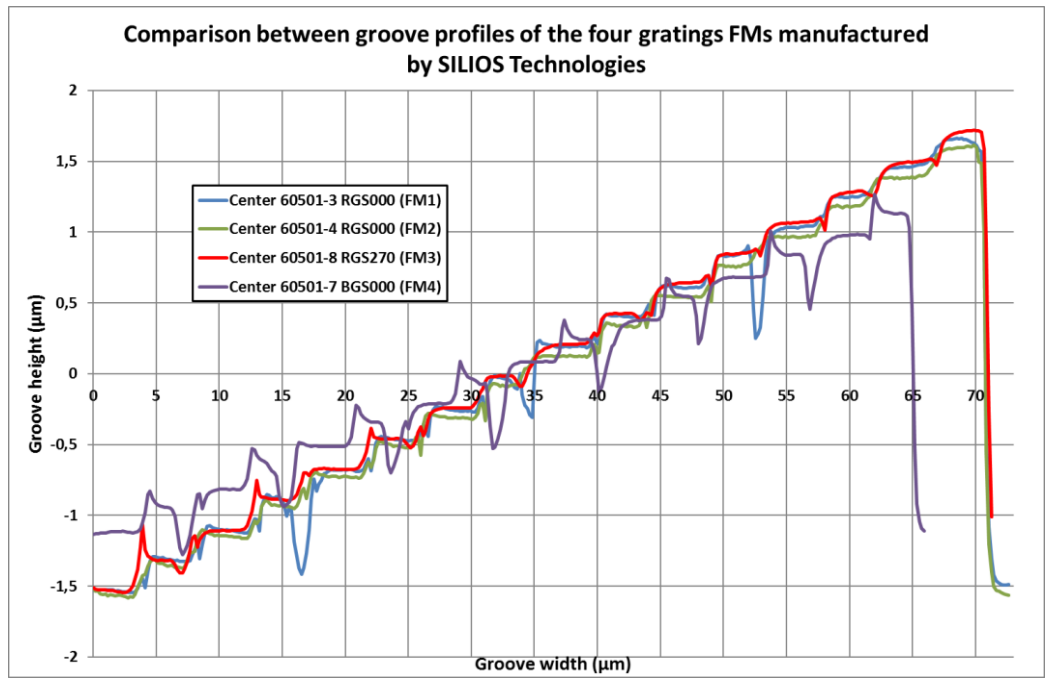

*Figure 4: Grating groove profile measured on the four FM grisms.*

### 3.2 Transmission measurement

The transmission performance of the NISP FM grisms has been measured at LAM onto the bandpass of each component in order 1 and order 0, and measured on the blocking band on samples with an accurate spectrophotometer. All components are within specifications in the bandpass as all transmission is better than 65% for each FM. In average on the four FM measurements, the mean efficiency of the FM is about 75% with a maximum transmission of 90%. This is a very good result for a component that combines a grating and a filter function. This is due to the high quality of the filter but also the high transmission done by the grating itself. In order 0, the average transmission is better than 1.4% for each model, which is also fine with respect to the calibration needs for NISP requiring more than 1%. Figure 5 presents the transmission curve of the NISP FM onto the bandpass done on the final components with filter and grating manufactured. We can remark that the transmission curves are very similar for the three red grisms, which demonstrate a very good reproducibility of the filter and the grating manufacturing process. One must note that the ripples that are seen on the curves are not due to the components but to the measurement set-up. Indeed, due to the dispersion of the grating and the focus introduced by the filter face, the measurement is done with the spectrophotometer associated with a specifically made fibered bench on small transmission bands of 100nm that samples the complete spectral band. The position of the output fiber going through the detector is optimized at the middle wavelength of each 100nm width band and the signal decreases at the edges of the spectral bandpass since the light beam go slightly out of this output fiber. The measurement of the grating transmission is particularly long and cannot be done in a one set measurement.

Figure 6 presents the transmission of the NISP FM on the full spectral band to show the blocking function of the filter. Concerning the blocking band, which is mainly due to the filter design, all components are within specifications except FM RGS270. For this component, there are some peaks on 2100-2500nm band that are fully acceptable for the project as they are in the extreme part of the blocking band and will not limit the performance of the instrument.

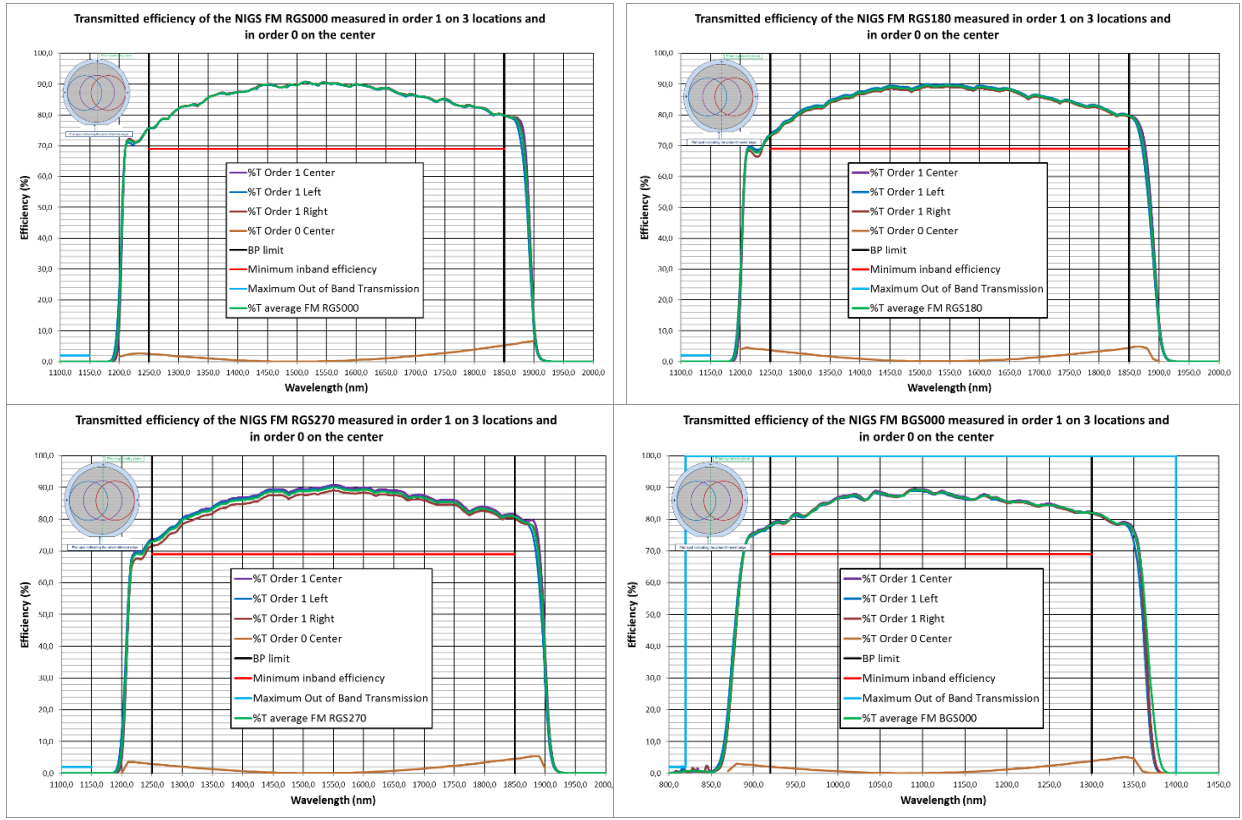

*Figure 5. Transmission curve of each NISP FM components onto the bandpass (1250-1850nm for "red" grisms, 950-1250nm for "blue" grism).*

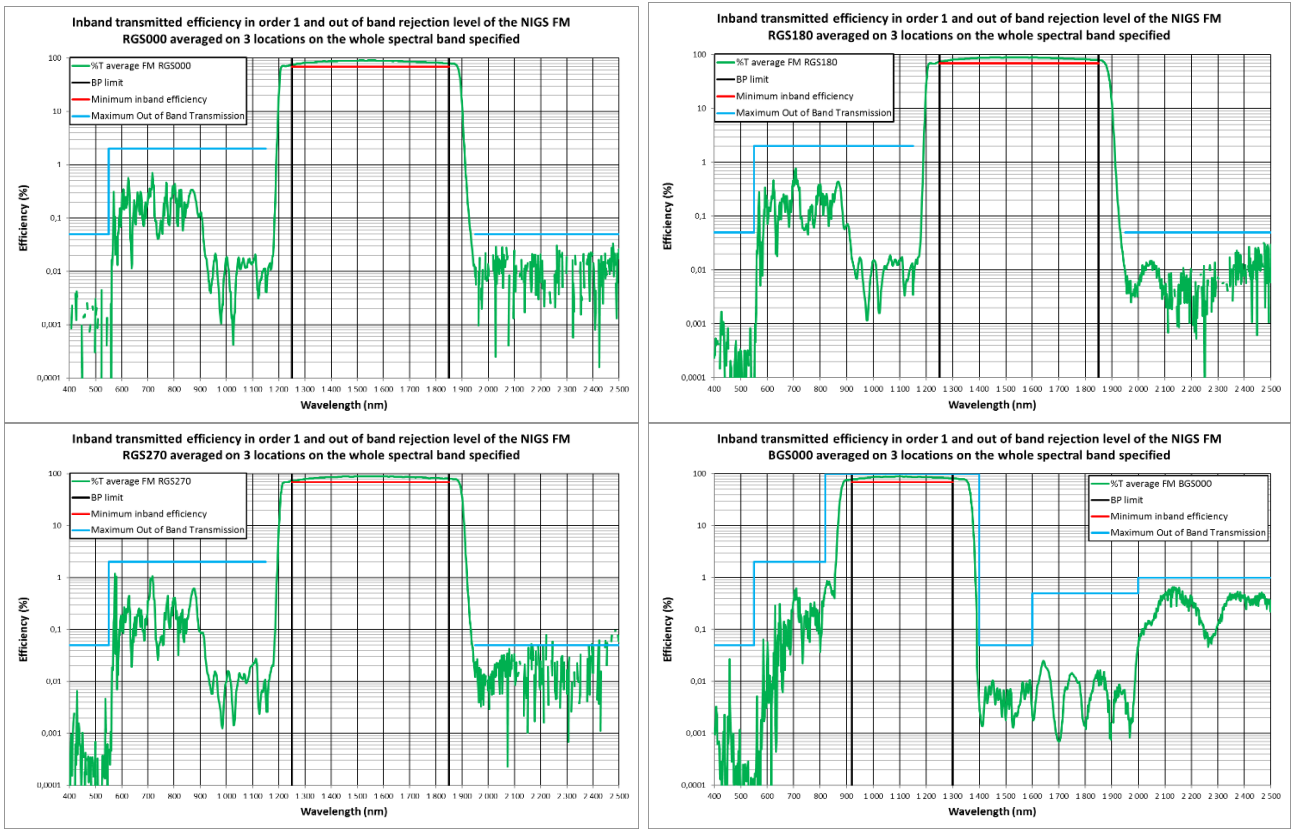

*Figure 6. Transmission curve of each NISP FM components onto the full spectral band to show the blocking band. Curves are shown in logarithmic scale.*

### 3.3 SFE performance

The other main characteristics to be validated for the grism FMs are the Surface Form Error (SFE) performance, obviously linked to the transmitted Wave Front Error (WFE) performance of the component. For the grisms, the specifications of the SFE and transmitted WFE have been split as the tolerance analysis of the NISP optical design has shown that many contributors are involved in the grisms performance and a classical characterization in transmission was difficult to implement. In addition, contrary to usual optical components, the nominal WFE of the grism is not close to zero i.e. a perfect WFE as the curved lines of the grating induce a chromatic wavefront correction. The wavefront measured with classical means is then compared to a nominal wavefront using a Zernike decomposition. This is then more difficult than measuring a small RMS value of the WFE. Finally, the focus of the grism introduced by the curvature radius of the filter surface is a critical specification. In fact, we need to have a co-focus better than 0.3fr relatively between the four grisms and the four filters of the NISP instrument as once everything integrated on the wheel and in the instrument, no focus adjustment capability is available.

Consequently, it was decided to separate the focus WFE budget from the rest of the WFE budget (without focus) as indicated in subsection 2.2 and in [6]. Another difficulty of the WFE measurement of the grisms was to distinguish effects from the manufacturing tolerances (surface polishing, error in grating manufacturing, etc.) from deformations effect due to the filter deposition, the gluing in the mechanical mount and the deformation of the whole component due to cryogenic environment. After FEM analysis of the component behavior under different load cases (filter deposition, gluing, interface defects, etc.), we have concluded that the same deformation is seen on each surface due to these load cases. The Zemax simulations of the transmitted WFE show a small effect on the transmitted WFE. We have then decided to characterize both surfaces of the grisms at each step of the manufacturing process to identify the impact of the different contributors and then to estimate with simulations the impact on the transmitted WFE.

The main issue for the grisms WFE measurement is that the grism does not transmit visible light once the filter is manufactured. Unfortunately, we have no facility at LAM to measure transmitted WFE in the NIR region. After discussion with the optical designer and the system engineer of NISP, it has been decided to rely on measurement only of SFE of each surface of the grism on 136mm clear aperture. The SFE and focus budgets have been distributed on each surface of the grism and also along the manufacturing process phases. This allows us to use only classical visible interferometer as described in [6] for the measurement of the SFE of the grisms. The performance and compliance of the FM with the WFE specifications are provided in Table 1. We can see that all "red" grisms reach the specifications defined by the project concerning the SFE and focus term after manufacturing. Only the BGS000 grism shows some non-compliances with the specification that has been accepted by the project as the optical performance of the blue channel will not be affected too much. One can note also that the curvature introduced by the filter deposition and the gluing is larger than expected for the FM RGS000 and FM RGS180. After discussion with the project, and analysis that this deformation has a small impact on the transmitted WFE, it has been agreed to accept the components with this non-compliance. Table 1 also indicates the type of verification used to validate the specification. The test set-up is fully described in [6] and uses a classical interferometer with a flat caliber to measure the grating face SFE and with a compensation lens to measure the filter face curvature. Only the compliance of the transmitted WFE is obtained thanks to analysis on Zemax software taking into account the measured SFE.

*Table 1. Summary of the WFE specifications and measurements obtained for all the grisms FM.*

| **Specification Description** | **Verification type** | **NI-GSU-FM-RGS000** | **NI-GSU-FM-RGS180** | **NI-GSU-FM-RGS270** | **NI-GSU-FM-BGS000** |
| --- | --- | --- | --- | --- | --- |
| Transmitted WFE < 20nm RMS @ 633mm (filter, gluing, mounting contribution) | Test & analysis | Compliant 12 nm RMS | Compliant 10 nm RMS | Compliant 12 nm RMS | Compliant 10 nm RMS |
| Focus error of the filter face of NIGSU shall be lower than +/- 0.5fringes from the nominal value | Test | Compliant 9633,23mm | Compliant 9631,06mm | Compliant 9272,43mm | Compliant 9908,27mm |
| Grating surface defocus < +/-0,5fr in order 1 | Test | Compliant 0,34fr | Compliant 0,05fr | Compliant 0,06fr | Non Compliant 0,61fr |
| Grating surface RMSi < 30nm RMS in order 1 | Test | Compliant 25,5nm RMS | Compliant 27 nm RMS | Compliant 27 nm RMS | Non Compliant 37,4nm RMS |
| Filter surface RMSi < 15 nm RMS | Test | Compliant 8,54 nm RMS | Compliant 10,5 nm RMS | Compliant 7 nm RMS | Compliant 3 nm RMS |
| Curvature of surfaces after gluing and filter deposition should not differ more than 5fr from the nominal value | Test | Non Compliant 5,52fr for grating 5,23fr for filter | Non Compliant 8,27fr for grating 8,4fr for filter | Compliant 4,63fr for grating 4,24fr for filter | Compliant 4,18fr for grating 4,42fr for filter |

The validation of the SFE performance of the grisms has shown that the grating manufacturing was very good and that the equation of the grating was properly engraved by the grating manufacturer. The specification of the grating surfaces takes into account both the manufacturing of the surface itself and the grating manufacturing error i.e. the difference with a perfect grating. The error due to the grating manufacturing for each FM is lower than 15 nm RMS. Figure 7 presents an example of the theoretical interferogram of the grating function of the RGS270 component compared with the measured interferogram on the FM component in order 5. We can see a very good similarity between both interferograms. After analysis, the difference between manufacturing and theory is lower than 15 nm RMS, focus error included. This is a very good result.

The method of SFE characterisation of the grisms has allowed following the evolution of the SFE of the component along the manufacturing process. In particular, it has permitted to identify the contribution of the filter manufacturing to the SFE of the component. As shown in Figure 8, the filter adds a large amount of focus to the SFE and also third order aberrations such as astigmatism and coma. Same aberrations have been added on each component, and measured on both grating and filter surfaces. The effect of these aberrations is small on the transmitted WFE. The impact on the focus deformation is not the same on each component: we have a bad reproducibility in terms of aberrations of the filter manufacturing as the defocus induced by the filter can vary between 4 and 8.5 fringes. In the design of the grisms, it is not possible to compensate for these aberrations as the filter is only deposited on one surface. It has also been chosen to not pre-compensate the optical surfaces, as the uncertainties on the defocus value were too large.

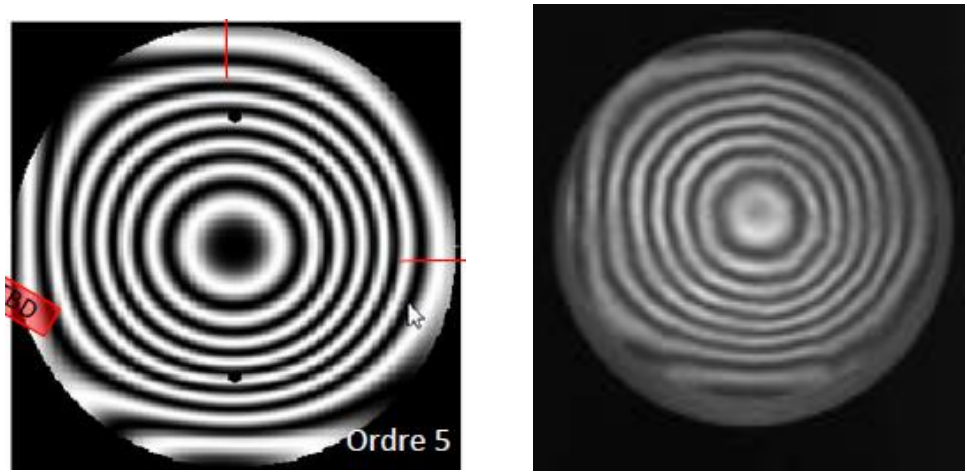

*Figure 7. Theoritical interferogram for the grating of RGS270 in order 5 (left) compared to the measured interferogram of the FM RGS270 (right).*

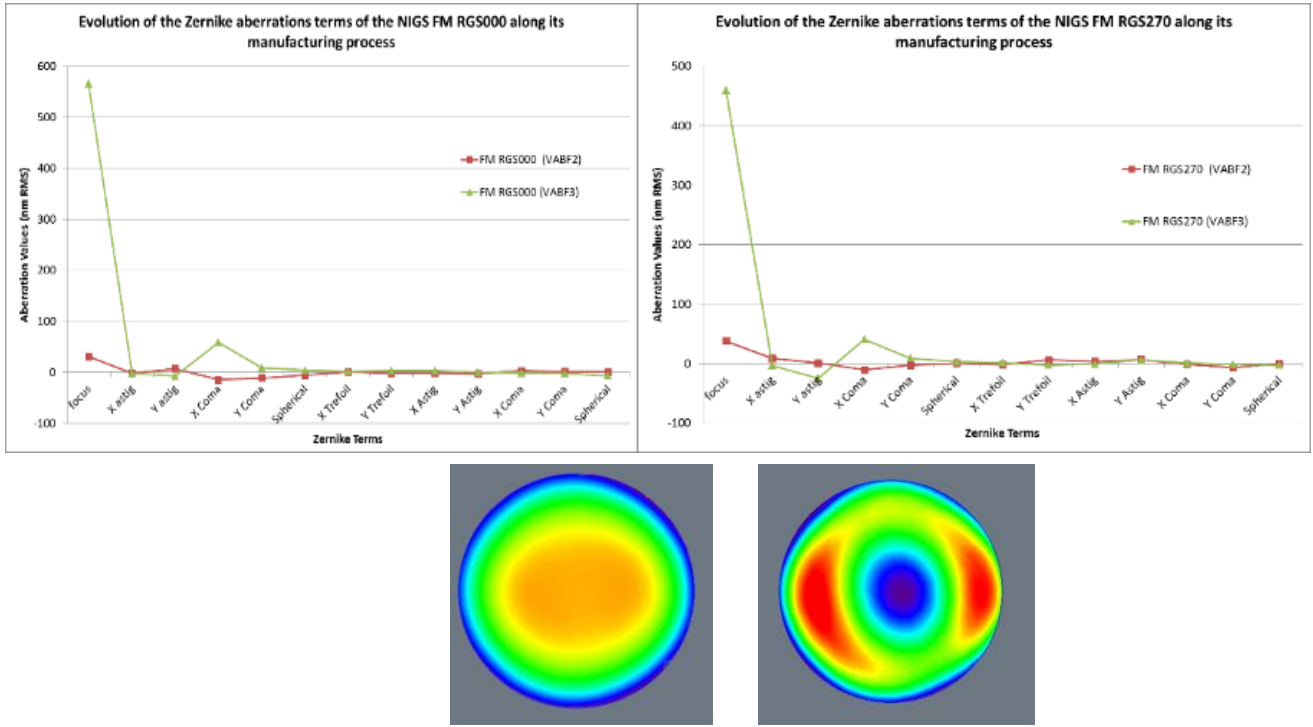

*Figure 8. Top: Evolution of the Zernike coefficient of the grating SFE before filter manufacturing (VABF2) and after filter manufacturing (VABF3). On the bottom, corresponding SFE maps are shown before filter manufacturing (left) and after filter deposition (right).*

### 3.4 Defocus measurement at cold temperature

In addition to the WFE measurement, it has been identified early in the project the need to verify that no defocus is introduced between room temperature and operational temperature i.e. 130K. In fact, all the SFE measurements are done at room temperature, but the specifications have to be verified at operational temperature. The most critical specification is the defocus. Due to time and cost of the development of an optical mean to validate SFE at operational temperature, it has been agreed to limit the measurement to a defocus measurement. The goal of the measurement is to see if any defocus between room temperature and operational temperature is present. Each grism FM has been measured thanks to a specific bench developed at LAM to measure the defocus at cold and shown in Figure 9. The principle of the bench is very simple:

- A source is illuminating the grism with a collimated light. The source is an Argon-Mercury lamp where we select with a filter the emission ray of 1540nm for the red grisms. The source is installed on a translation stage through the optical axis in order to perform the defocus on the source itself;
- The grism is installed in a vacuum chamber: two windows of 90mm aperture on the two doors of the chamber are present to allow light to pass through the component and to measure the defocus of the component in transmission. The filter surface is in the direction of the source;
- The light passing through the chamber and the grism is focalised onto an IR detector to measure the PSF. The IR detector is a commercial camera from Xenics with an InGas detector, cooled down with a Peltier system at -80°C. The camera has not a very good quality and it has a certain level of noise that is why we are not doing image performance characterisation with the set-up.

The defocus measurement consists first to acquire a through focus curve of the system at ambient temperature and vacuum pressure to know the position of the best focus in ambient conditions. This is the "zero" point of our set-up to which we will compare the other measurements. Then the chamber is cooled down to reach the operational temperature of the component i.e. 130K on the optical part and a new measurement of the best focus curved is done and compared to the measurement done at ambient conditions. The goal of the test is to verify that a difference of focus lower than 0.1mm is found as a larger defocus will have an impact on the optical quality for NISP. Figure 10 shows the results obtained for the FM RGS000. We can see that the best focus position is similar between ambient and operational temperature and lower than 0.1mm after fitting of the data with a Gaussian profile. Figure 11 shows the PSF obtained with the test set-up. One must note that these PSF are not "NISP" like PSF as the sampling of the PSF is specific to our bench and obviously, the image quality of the system is much worse than the one from NISP instrument. Nevertheless these PSF are interesting because they allow seeing that no difference is observed at ambient and operational temperatures. It is also very interesting to look at the grism effect on the full spectrum image: we clearly see the defocus of the different emission rays of the lamp which is the main function of the grism of NISP that compensates for the NISP detector tilt. If we go through focus, we are focusing one after the other the emission rays of the lamp.

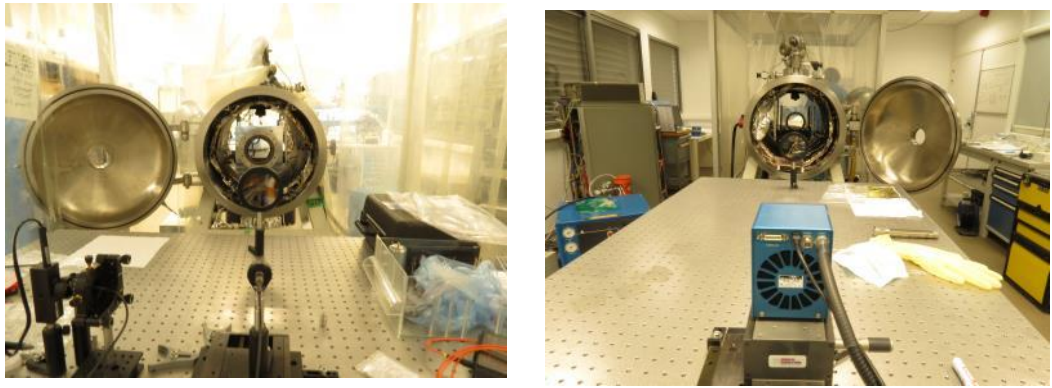

*Figure 9. Picture of the defocus bench seen from the light source side (left) and from the camera side (right). In the chamber, the grism is seen while the doors of the chamber are opened.*

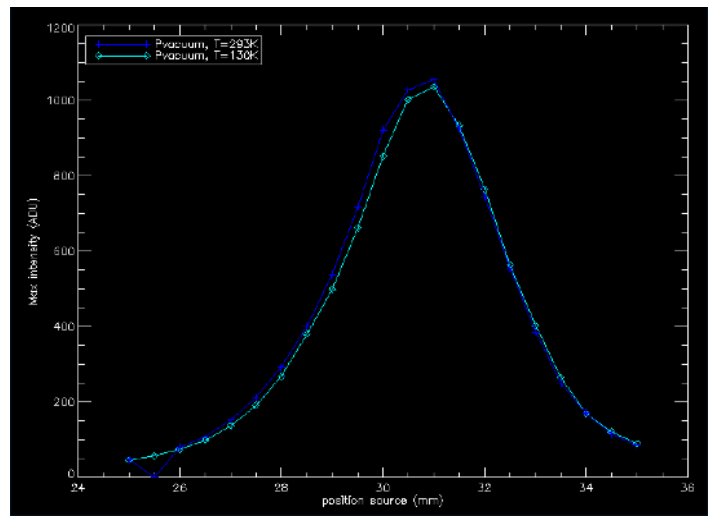

*Figure 10. Defocus curve obtained for FM RGS000 between ambient temperature and operational temperature.*

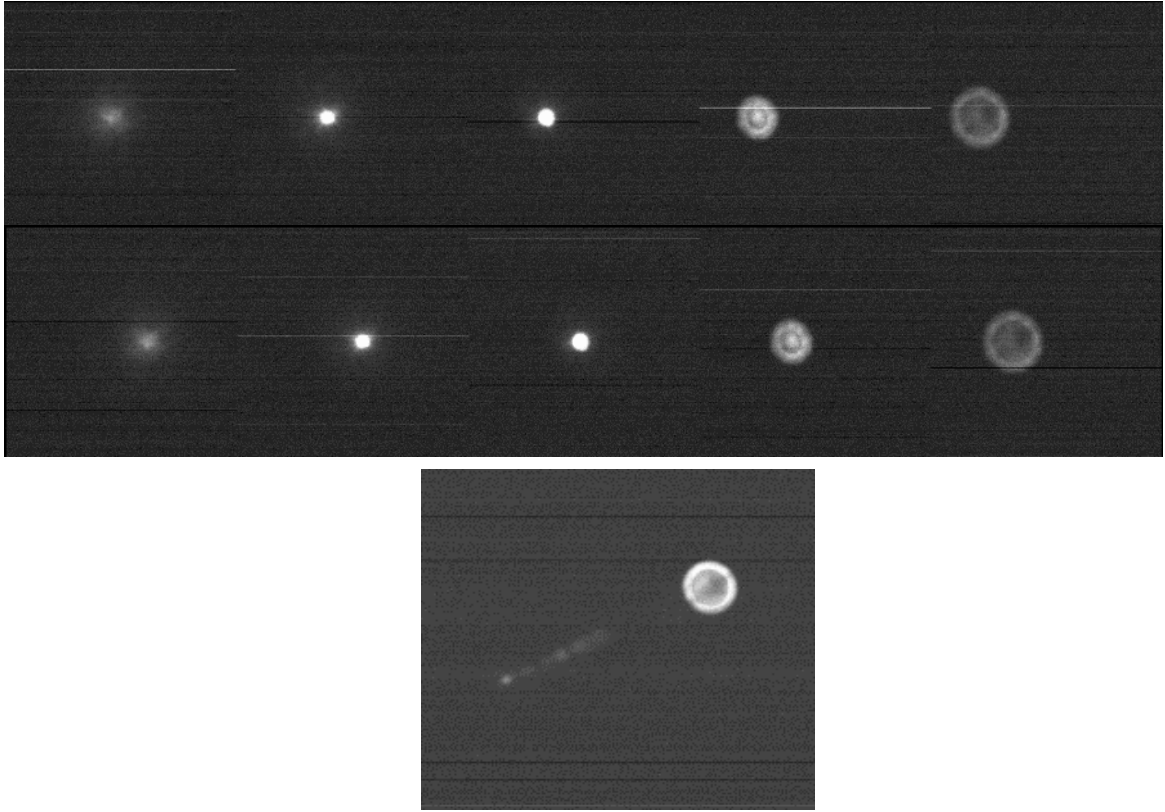

*Figure 11. Top: PSF obtained at different focus positions of the source at ambient temperature (top) and operational temperature (middle) at 1540nm. The middle PSF is close to the best focus position. Bottom: image of the PSF obtained with the HgAg lamp without filter: all emission rays of the lamp are seen. We clearly see the defocus induced by grating from the left to the right.*

## 4. MECHANICAL PERFORMANCE OF NISP FM

In addition to the optical performance, we have to demonstrate the resistance of the grisms FM to the level of vibration defined for acceptance of the component. After successful campaign done on the engineering model [5], we have performed acceptance vibrations tests with the following specifications

- Sine acceptance levels of 30.5g on X axis, 22.3g on Y and Z axes;
- Random test of 11.6g RMS in X, 10.6g RMS in Y and 9.9g RMS in Z with notches at the resonant frequencies of the grism.

All acceptance tests were completed at LAM in ISO5 environment for all FM components at the needed specifications as shown in Figure 13. Very good correlations were observed between the tests results and the FEA for all sensors and all directions for all components. Some examples of the results obtained are shown in Figure 12. What was interesting to see is that the behavior of each component was similar also one respect to the others. There was no specific response of one of the component compared to the other. This is due to the fact that the geometry, mass and dimension of all components were similar. A very good reproducibility of the mechanical manufacturing has been reached. The success of this test has allowed demonstrating the compliance of the design mount with the need for EUCLID mission and the performance of the manufactured components for NISP FM.

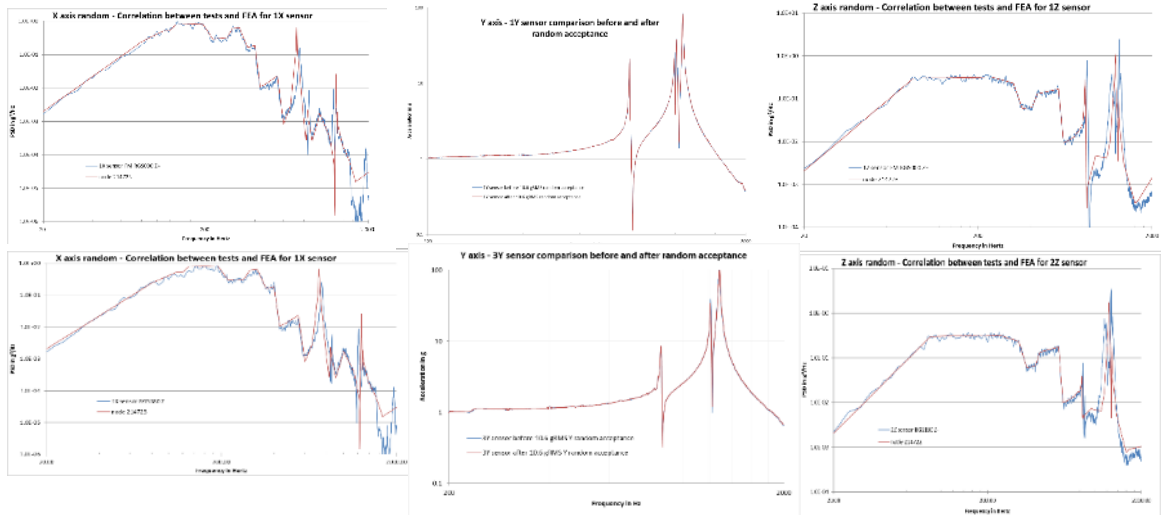

*Figure 12. Comparison between FEA simulation and measured response for FM RGS180 (top) and FM RGS000 (bottom) respectively for X, Y and Z axis random test.*

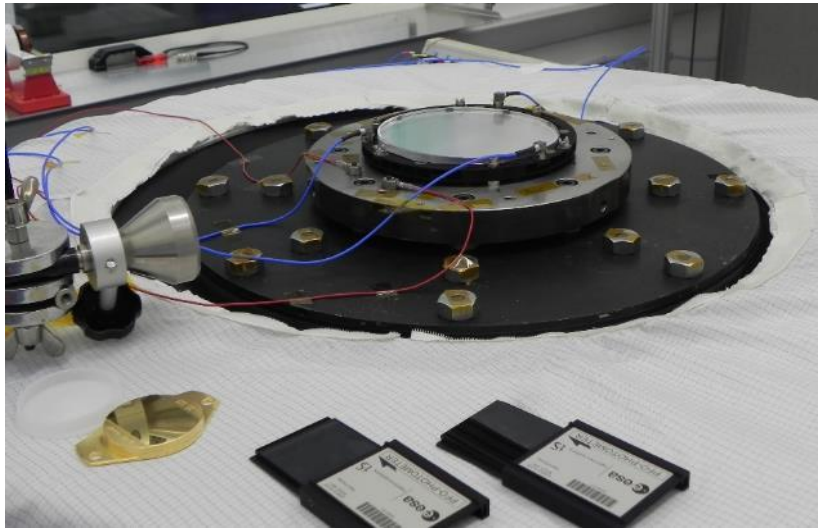

*Figure 13. Picture of one of the NIGS FM during vibration runs in ISO5. A blanket covers the shaker to avoid over contamination while an ISO5 laminar flux is run during the test.*

## 5. CONCLUSION

We have presented in this paper the overall performance of the NISP flight models for EUCLID mission. The grisms manufactured, integrated and tested under the LAM responsibility have shown a very well compliance with the specifications of the project. The analyses of the tests results have shown that the grisms flight models for NISP are within specifications with an efficiency better than 70% on the spectral bandpass and a wavefront error on surfaces better than 30nm RMS. The components have withstood vibration qualification level up to 11.6g RMS in random test and vacuum cryogenics test down to 130K with measurement of optical quality in transmission.

The EUCLID grisms flight models have been delivered to the NISP grism wheel in November 2017 and are now fully integrated onto the wheel that will be delivered to NISP instrument in June 2018. Assembly and alignment of NISP instrument has already started and shall be finished by end of 2018. The grism FM will be tested and validated in NISP in early 2019 to demonstrate the full performance of the spectroscopic mode of NISP.

# REFERENCES

[1]G. Racca; R. Laureijs; L. Stagnaro; J.-C. Salvignol, et al, "The Euclid mission design," *Proc. SPIE* 9904, Space Telescopes and Instrumentation 2016: Optical, Infrared, and Millimeter Wave, 99040-23 (2016)

[2] T. Maciaszek; et al., "Euclid near infrared spectrophotometer instrument concept and first test results at the end of phase C," *Proc. SPIE* 9904, Space Telescopes and Instrumentation 2016: Optical, Infrared, and Millimeter Wave, 99040-18 (2016)

[3] M. Cropper; S. Pottinger; S. Niemi; J. Denniston; et al., "VIS: the visible imager for Euclid," *Proc. SPIE* 9904, Space Telescopes and Instrumentation 2016: Optical, Infrared, and Millimeter Wave, 99040-16 (2016)

[4] A. Costille; A. Caillat; C. Rossin; S. Pascal; B. Foulon; P. Sanchez; S. Vives, "Final design and choice for EUCLID NISP grism," *Proc. SPIE* 9912, Advances in Optical and Mechanical Technologies for Telescopes and. Instrumentation II 2016, 99122C (2016)

[5] C. Rossin; A. Costille, A. Caillat, S. Pascal, P. Sanchez, et Al. "Final design of the Grism cryogenic mount for the Euclid-NISP mission", *Proc. SPIE* 9912, Advances in Optical and Mechanical Technologies for Telescopes and. Instrumentation II 2016, paper 991261 (2016)

[6] A. Caillat; A. Costille; S. Pascal; S. Vives; C. Rossin; P. Sanchez; B. Foulon, " Optical verification tests of the NISP/Euclid grism qualification model," *Proc. SPIE* 9904, Space telescopes and Instrumentation 2016: Optical, Infrared, and Millimeter Wave, 99040R (2016)

[7] TRIOPTICS, http://www.trioptics.com/

[8] SILIOS Technologies, http://www.silios.com/

[9] WINLIGHT Optics, http://www.winlight-system.com/

[10] Optics Balzers Jena, http://www.opticsbalzers.com/

[11] A. Caillat; S. Pascal; S. Tisserand; K. Dohlen; R. Grange, et al., " Bulk silica transmission grating made by reactive ion etching for NIR space instruments ", *Proc. SPIE* 9151, Advances in Optical and Mechanical Technologies for Telescopes and Instrumentation, 91511F (2014)